\documentclass{ws-procs9x6}
\newcommand{\fcdz}       {\mbox{$f(c \rightarrow D^0)$}}
\newcommand{\fcdc}       {\mbox{$f(c \rightarrow D^+)$}}
\newcommand{\fcdss}      {\mbox{$f(c \rightarrow D_s^+)$}}
\newcommand{\fclc}       {\mbox{$f(c \rightarrow \Lambda_c^+)$}}
\newcommand{\fcds}       {\mbox{$f(c \rightarrow D^{\ast +})$}}
\newcommand{\dspm}       {\mbox{$D^{\ast \pm}$}}

\begin{document}

\title{Experimental results on heavy quark fragmentation
\footnote{\uppercase{P}articipation in \uppercase{DIS} 2006 was supported by the local organising committee and grant 06-02-26609-z of the
\uppercase{R}ussian \uppercase{F}oundation for \uppercase{B}asic
\uppercase{R}esearch.}}

\author{L.~K.~GLADILIN}

\address{Skobeltsyn Institute of Nuclear Physics, Moscow State University\\
Vorob'evy Gory, Moscow, RU-119992, Russia\\
E-mail: gladilin@sinp.msu.ru }

\maketitle

\abstracts{
Experimental results on $c$- and $b$-quark fragmentation are reviewed.
The discussion is concentrated on measurements of heavy-quark fragmentation
functions
and 
fragmentation fractions.
Measurements of various heavy-quark fragmentation ratios are also discussed.
The experimental results are compared with theoretical expectations and
model predictions.
}

\section{Introduction}

The initial stage of charm/bottom quark fragmentation
can be described by perturbative QCD (pQCD) calculations~\cite{oleari_dis06}.
A non-perturbative (NP) parameterisation is needed to describe
the final heavy-quark transformation to
a particular charmed or bottom hadron.
Such parameterisation can include
effects producing by the excited states decaying to a given hadron.
The NP fragmentation parameterisation can be splited in two
parts: fragmentation function and fragmentation fraction.
Fragmentation functions are used to parameterise the transfer
of the quark's energy to a given meson; they can be different
for different pQCD calculations used to describe the initial
fragmentation.
Fragmentation fractions
are the fractions of $c$/$b$ quarks hadronising as a particular
charmed/bottom hadron; they are expected to be
universal for all pQCD calculations.

Measurements of the heavy quark fragmentation allow testing
pQCD calculations and extracting fragmentation functions
and fractions.
A deeper phenomenological understanding of the heavy quark fragmentation can
be obtained by measuring
various heavy-quark fragmentation ratios.
In particular, we will discuss
the ratio of neutral and charged $D$/$B$ meson production rates, $R_{u/d}$,
the strangeness-suppression factor, $\gamma_s$, and the fraction
of $D$/$B$ mesons produced in a vector state, $P_{\rm v}$.


\section{Bottom quark fragmentation}

The $b$-quark fragmentation function was measured at
LEP~\cite{aleph_ff,opal_ff,delphi_ff}
and SLD~\cite{sld_ff}.
The measured spectra were compared with
predictions of
the leading-logarithmic (LL) JETSET~7.4~\cite{pythia} Monte Carlo (MC)
using different  parametrisations
for the fragmentation function.
The best description of the data with a parametrisation
with one free parameter was obtained using the parametrisation
of Kartvelishvili et al.~\cite{kartve}.
The Bowler~\cite{bowler} and symmetric LUND~\cite{lund}
parametrisations with two free parameters provided a better
data description.
The Peterson~\cite{peterson} and Collins-Spiller~\cite{collins}
parameterisations, and the HERWIG cluster model~\cite{herwig} predictions
were found to be too broad to describe the data.
The $b$-quark fragmentation function measurements were also used for fitting
the NP parametrisation with the next-to-leading-order (NLO)
calculations~\cite{cacciari_frag,kniehl_bfrag}.

The $b$-quark fragmentation fractions
were obtained by combining
of all published LEP and CDF results on production of
the weakly decaying $B$ hadrons with measurements of the time-integrated
mixing probabilities~\cite{hfag,pdg04}.
The isospin invariance, i.e. $R_{u/d}=1$, was assumed in this procedure.
Using the measured fragmentation fractions, the strangeness-suppression factor
for bottom mesons is
$$\gamma_s=2 f(\bar{b}\rightarrow B^0_s)/[f(\bar{b}\rightarrow B^0) + f(\bar{b}\rightarrow B^+)]=0.27\pm0.03.$$
Thus, bottom-strange meson production is suppressed by a factor $\approx3.7$.
The combined LEP value for the fraction
of $B$ mesons, produced in a vector state, is $P_{\rm v}=0.75\pm0.04$~\cite{pdg04}, that is in perfect agreement with
the naive spin counting expectation (0.75).

\section{Charm quark fragmentation}

The $c$-quark fragmentation function has been recently measured with
high precision by the CLEO~\cite{cleo_ff} and
BELLE~\cite{belle_ff} collaborations.
The data comparison with
the JETSET MC predictions revealed the same picture
as for the $b$-quark fragmentation.
The best description of the data was obtained
using the Bowler parametrisation with two free parameters,
and the parametrisation
of Kartvelishvili et al. with one free parameter.

A discrepancy between the NP parametrisations
obtained with
the CLEO/BELLE data and earlier
ALEPH measurement~\cite{aleph_cc_ff}
has been observed
using the NLO initial conditions, next-to-leading
logarithmic (NLL) evolution, NLO coefficient functions and NLL Sudakov
resummation~\cite{oleari_dis06,cacciari_frag}.
The difference, which was attributed
to the evolution between the $\Upsilon(4S)$ and $Z^0$ energies,
results in an additional uncertainty in predictions for $\dspm$
hadroproduction of the order $20\%$.
To reduce the uncertainty 
direct measurements of the charm fragmentation function
at hadronic machines would be useful. Such measurements were
already performed in $ep$ interactions at HERA by
the ZEUS~\cite{zeus_ffunction} and H1~\cite{h1_ffunction} collaborations;
their results were found to be 
in qualitative agreement with
those obtained in $e^+e^-$ annihilations. 
\begin{table}[ph]
\tbl{
The fractions of $c$ quarks hadronising as a particular charm hadron,
$f(c \rightarrow D, \Lambda_c)$.
The fractions are shown for the $D^+$, $D^0$, $D^+_s$
and $\Lambda_c^+$ charm ground states and for the
$D^{*+}$ state.
}
{\footnotesize
\begin{tabular}{|c|c|c|c|} \hline
& & Combined & \\
& ZEUS ($\gamma p$)~\cite{zeus_php_fb} & $e^+e^-$ data~\cite{hep-ex-9912064} & H1 (DIS)~\cite{h1_dis_fb} \\
\hline
\hline
&\phantom{~~~~~~~~~} stat.\phantom{~~} syst.\phantom{~} br.
&\phantom{~~~~}stat.$\oplus\,$syst.\phantom{~} br.
&\phantom{~~~~~~~~~~}total \\
\hline
$\fcdc$ & $0.217 \pm 0.014\phantom{~}^{+0.013\, +0.014}_{-0.005\, -0.016}$  &
$0.226\phantom{~} \pm 0.010\phantom{~~} ^{+0.016}_{-0.014}$ & $0.203 \pm 0.026$ \\
\hline
$\fcdz$ & $0.523 \pm 0.021\phantom{~}^{+0.018\, +0.022}_{-0.017\, -0.032}$  &
$0.557\phantom{~} \pm 0.023\phantom{~~} ^{+0.014}_{-0.013}$ & $0.560 \pm 0.046$ \\
\hline
$\fcdss$ & $0.095 \pm 0.008\phantom{~}^{+0.005\, +0.026}_{-0.005\, -0.017}$ &
$0.101\phantom{~} \pm 0.009\phantom{~~} ^{+0.034}_{-0.020}$ & $0.151 \pm 0.055$ \\
\hline
$\fclc$ & $0.144 \pm 0.022\phantom{~}^{+0.013\, +0.037}_{-0.022\, -0.025}$ &
$0.076\phantom{~} \pm 0.007\phantom{~~} ^{+0.027}_{-0.016}$ & \\
\hline
\hline
$\fcds$ & $0.200 \pm 0.009\phantom{~}^{+0.008\, +0.008}_{-0.006\, -0.012}$ &
$0.238\phantom{~} \pm 0.007\phantom{~~}^{+0.003}_{-0.003}$ & $0.263 \pm 0.032$ \\
\hline
\end{tabular}\label{table1} }
\vspace*{-13pt}
\end{table}

Table~1 compares the $c$-quark fragmentation fractions measured
in $ep$ interactions at HERA by
the ZEUS~\cite{zeus_php_fb} and H1~\cite{h1_dis_fb} collaborations
with those obtained in $e^+e^-$ annihilations.
The latter values were compiled previously~\cite{hep-ex-9912064}
and updated with the recent branching ratio values~\cite{pdg04}.
The measurements performed in $e^+e^-$ and $ep$ interactions are consistent.
Measurements of the $R_{u/d}$ value in charm fragmentaion confirmed
isospin invariance~\cite{zeus_php_fb,h1_dis_fb,hep-ex-9912064,zeus_dis_fb}.
Meausurements of the strangeness-suppression factor
in charm fragmentation showed
that charmed-strange meson production is suppressed
by a factor $\approx3.9$
(similar to the suppression in bottom fragmentation).
The fraction of charged $D$ mesons produced in a vector state,
$P_{\rm v}^d$,
was found to be
$\approx 0.6$\cite{zeus_php_fb,h1_dis_fb,hep-ex-9912064,zeus_dis_fb}
in both $e^+e^-$ and $ep$ interactions. The value is significantly smaller
than that obtained in bottom fragmentation and does not agree with
the naive spin counting expectation ($0.75$).

\section{Summary}

The $b$-qaurk fragmentation function and fractions were measured
in  $e^+e^-$ annihilations, while the $c$-quark fragmenation was studied
in both $e^+e^-$ and $ep$ interactions.
Comparison of the charm fragmentation characteristics, obtained
in $e^+e^-$ and $ep$ interactions, generally supports the hypothesis that
fragmentation proceeds independently of the hard sub-process.

The fraction of charged $D$ mesons produced in a vector state,
$P_{\rm v}^d$, in charm fragmentation was found to be
$\approx 0.6$
in both $e^+e^-$ and $ep$ interactions. The value is significantly smaller
than that obtained in bottom fragmentation and does not agree with
the naive spin counting expectation ($0.75$).

\end{document}